\documentclass[aps,pra,floatfix,twocolumn,showpacs,preprintnumbers,amsmath,amssymb,a4paper,superscriptaddress]{revtex4-1}
\usepackage{braket}
\usepackage{bbold}
\usepackage{dsfont}
\usepackage{amsmath}
\usepackage{amssymb}
\usepackage{amsfonts}
\usepackage{mathtools}
\usepackage{graphicx}
\usepackage{dcolumn}
\usepackage{bm}
\usepackage{multirow}
\usepackage{leftidx}
\usepackage{color}
\usepackage{MnSymbol}
\usepackage[mathscr]{eucal}
\usepackage[german,english]{babel}
\usepackage{array}
\usepackage{pifont}%
\usepackage[capitalize]{cleveref}
\usepackage[utf8]{inputenc}
\usepackage{natbib}

\begin{document}

\title{Asymmetric lineshapes of Efimov resonances in mass-imbalanced ultracold gases}
\author{P. Giannakeas}
\email{pgiannak@pks.mpg.de}
\affiliation{Max-Planck-Institut f\"ur Physik komplexer Systeme, N\"othnitzer Str.\ 38, D-01187 Dresden, Germany }

 \author{Chris H. Greene}

 \affiliation{Department of Physics and Astronomy, Purdue University, West Lafayette, Indiana 47907, USA}
 \affiliation{Purdue Quantum Science and Engineering Institute, Purdue University, West Lafayette, Indiana 47907, USA}
 \date{\today}

\begin{abstract}
The resonant profile of the rate coefficient for three-body recombination into a shallow dimer is investigated for mass-imbalanced systems.
In the low-energy limit, three atoms collide with zero-range interactions, in a regime where the scattering lengths of the heavy-heavy and the heavy-light subsystems are positive and negative, respectively.
For this physical system, the adiabatic hyperspherical representation is combined with a fully semi-classical method and we show that the shallow dimer recombination spectra display an asymmetric lineshape that originates from the coexistence of Efimov resonances with St\"uckelberg interference minima.
These asymmetric lineshapes are quantified utilizing the Fano profile formula.
In particular, a closed form expression is derived that describes the width of the corresponding Efimov resonances and the Fano lineshape asymmetry parameter $q$.
The profile of Efimov resonances exhibits a $q-$reversal effect as the inter- and intra-species scattering lengths vary.
In the case of a diverging asymmetry parameter, i.e. $|q|\to \infty$, we show that the Efimov resonances possess zero width and are fully decoupled from the three-body and atom-dimer continua,  and the corresponding Efimov metastable states behave as bound levels.
\end{abstract}

\maketitle
\section{Introduction}
The Efimov effect is one of the most counter-intuitive phenomena in few-body physics where an infinity of three-body bound states is formed even when the scattering length of the two-body subsystems is negative \cite{nielsen_three-body_2001,greene_review_2017,naidonEfimovPhysicsReview2017,dincaoFewbodyPhysicsResonantly2018}.
This phenomenon was theoretically predicted by V. Efimov to occur for 3 equal mass particles that interact via zero-range potentials, with trimer binding energies that scale geometrically \cite{v_efimov_hard-core_1970}.
The existence of these exotic trimer states was experimentally confirmed by Kraemer {\it et al.} in an ultracold gas of $\rm{Cs}$ atoms \cite{kraemer_evidence_2006}.
This suggested new possibilities for theoretical and experimental investigations \cite{nielsen_three-body_2001,riisager_nuclear_1994,braaten_universality_2006,rittenhouse_hyperspherical_2011,blume_few-body_2012,wang_chapter_2013,wang2015few,naidon_review_2017,greene_review_2017} to address various physical aspects of the Efimov states, such as the discrete scale invariance of the trimer binding energies \cite{huang_observation_2014} or the sensitivity of the ground state energy on the short-range physics.
In particular, the latter stems from the fact that within the zero-range model the trimer spectrum is unbound from below due to {\it Thomas collapse} \cite{thomas_interaction_1935-1} and an auxiliary parameter, i.e. three-body parameter, was introduced in order to specify the ground state energy rendering the entire spectrum system dependent \cite{v_efimov_hard-core_1970}.
However, experimental and theoretical advances demonstrated that for ultracold atoms the Efimov spectrum exhibits a certain class of universality, i.e. van der Waals universality \cite{wang_universal_2014-1,roy_prl_2013,wang_origin_2012,gross_observation_2009,naidon_microscopic_2014,naidon_physical_2014,ferlaino_forty_2010,giannakeasfewbody2016,mestromEfimovVanWaals2017,mestromVanWaalsUniversality2020,kunitski_observation_2015}.
Namely, it was shown that the lowest Efimov state appears at scattering lengths $a^{(1)}_-\approx -10 \ell_{\rm{vdW}}$ with $\ell_{\rm{vdW}}$ being the length scale of van der Waals interactions between two neutral atoms.

Mass-imbalanced ultracold gases are an ideal platform to explore more deeply the idiosyncrasies of Efimov spectra.
In particular, three-body collisions of ultracold atoms with unequal masses offer more favorable experimental conditions that enable observation of multiple successive trimer states and measurement of their geometrical energy scaling, i.e. the smoking gun of Efimov physics \cite{pires_observation_2014prl,ulmanis_universality_2015,tung_geometric_2014-1}.
Apart from that, mass-imbalanced ensembles offer a large parameter space, such as the particles' mass-ratio, the sign and magnitude of the inter- and intra-species scattering lengths, which provide fertile ground to investigate pristine attributes of the Efimov states.
Specifically, theoretical and experimental efforts have mapped out a large portion of the parameter space addressing the underlying physics of recombination processes in heavy-heavy-light (HHL) systems \cite{pires_observation_2014prl, ulmanis_universality_2015,hafner2017role,petrov_three-body_2015,zhaopra2019,sun2021efimov}.
The particular case which stands out corresponds to HHL systems that possess inter- and intraspecies scattering lengths of opposite sign, i.e. $a_{HL}<0$ and $a_{HH}>0$ respectively.
For example, in the experimental works of Refs.\cite{ulmanis_heteronuclear_2016,hafner2017role}, it was demonstrated in the regime of broad Fano-Feshbach resonances \cite{chin_feshbach_2010} that the lowest Efimov state is in good agreement with the predictions of the universal zero-range and van der Waals theory.
However, subsequent experimental investigations show that deviations from the universal theory are more pronounced for narrow Fano-Feshbach resonances \cite{johansen_testing_2017}.
Furthermore, within the zero-range theory Ref.\cite{giannakeasprl2018} illustrated that the diabaticity of the three-body collisions imposes additional limitations on the universal properties of Efimov spectrum, where mostly adiabatic collisions yield trimer states independent of the three-body parameter, as was pointed out in the case of Refs.\cite{ulmanis_heteronuclear_2016,hafner2017role}.

Additionally, Ref.\cite{giannakeasprl2018} showed that three-body recombination into a shallow heavy-heavy dimer possesses a unique property that only mass-imbalanced systems exhibit, namely the co-existence of Efimov resonances with St\"uckelberg suppression effects in the same range of scattering lengths.
In this work, we further study this particular attribute of HHL systems and demonstrate that the corresponding Efimov resonances in the recombination rate coefficient plotted versus scattering length can display an asymmetric profile, which can be quantified by the Fano profile formula.
In particular, our analysis employs the adiabatic hyperspherical framework for zero-range two-body interactions which is combined with a fully semi-classical theory \cite{giannakeasprl2018}.
Also, a simplified version of the semi-classical approach is shown where the lowest hyperspherical curves are approximated by universal potential tails at large hyperradii, as in Ref.\cite{dincao_scattering_2005}.
This permits us to derive closed form relations for the $S$-matrix elements which are expressed in terms of the width of the Efimov resonance and Fano's lineshape asymmetry parameter $q$.
As an example, the asymmetric profiles of the Efimov resonances in the recombination coefficient of $^6\rm{Li}-^{133}\rm{Cs}-^{133}\rm{Cs}$ and $^6\rm{Li}-^{87}\rm{Rb}-^{87}\rm{Rb}$ are analyzed, both of which showcase a $q$-reversal phenomenon as function of the inter- and intraspecies scattering length ratio.
Furthermore, we observe that for a diverging $q$ parameter the Efimov resonances behave as bound states that are embedded in the continuum \cite{hsuBoundStatesContinuum2016}.
This occurs since the decay width of the resonances vanishes as $|q|\to \infty$ and the corresponding Efimovian quasi-bound states decouple from the three-body and atom-dimer continua.

The structure of this work is as follows: In \cref{sec:sec1} the Hamiltonian of the three-body system and the parameters of interest are given.
\cref{sec:sec1a,sec:sec1b}  provide a detailed review of the methods that are employed in our analysis.
More specifically, \cref{sec:sec1a} discusses the adiabatic hyperspherical representation and the fully semi-classical treatment of the coupled hyperradial equations.
In \cref{sec:sec1b} a simplified version of the semi-classical theory is given that permits us to express the $S-$ matrix elements of recombination processes into shallow dimers in terms of the inter- and intra-species scattering lengths.
Finally, \cref{sec:sec2} focuses on the asymmetric profile of Efimov resonances in the spectrum of the three-body recombination coefficient for HHL systems.

\section{General considerations and methods} \label{sec:sec1}
Consider a three-body system that consists of two heavy (H) alkali atoms and a light (L) one at low-energies. 
The particles mutually interact through $s$-wave pairwise interactions that are modeled via Fermi-Huang's zero-range pseudopotential.
Our greatest interest here is in the regime where the mass-imbalanced system can recombine into a shallow heavy-heavy dimer with a recoiling light atom.
This scenario arises for inter- and intraspecies interactions of opposite sign meaning that the scattering length between a heavy-light or heavy-heavy pair of particles is $a_{HL}<0$ or $a_{HH}>0$, respectively.
Furthermore, relaxation and recombination processes into deep dimer channels will be neglected, and we will focus on the physics that arises due to energies near the break-up threshold, i.e. the zero-energy limit.
For this purpose, we focus on the two lowest potential curves of HHL systems, which suffice to describe three-body recombination processes into shallow dimers as was shown in Ref.\cite{giannakeasprl2018} permitting the derivation of closed form expressions for the $S-$matrix.

\subsection{The adiabatic hyperspherical representation and the semi-classical approach} \label{sec:sec1a}
The total three-body Hamiltonian for the HHL system of interest is given by the following expression:
\begin{equation}
\begin{aligned}
    H_{\rm{tot}}= \sum_{i=1}^3 -\frac{\hbar^2}{2 m_i} \nabla_i^2 + \sum_{i>j} V_{ij}(\boldsymbol{r}_{ij}),~~\rm{with}\\
    V_{ij}(\boldsymbol{r}_{ij})=\frac{4\pi \hbar^2a_{ij}}{2 \mu_{ij}} \delta (\boldsymbol{r}_{ij}) \partial_{r_{ij}}[r_{ij} \times],
\end{aligned}
    \label{eq:eq1}
\end{equation}
where $V_{ij}$ represents the Fermi-Huang pseudopotential. $a_{ij}$ and $\mu_{ij}$ refer to the scattering length and two-body reduced mass of the $ij$-pair of particles, respectively. $\nabla^2_i$ denotes the Laplacian for $\boldsymbol{r}_i$, and $m_i$ indicates the mass of the $i$-th particle.
Note that the scattering lengths $a_{ij}$ between the atoms are chosen to be larger than any other length scale of the system permitting us to focus on the universal characteristics of the three-body system under consideration.

Utilizing the Jacobi vector choice of Ref. \cite{rittenhouse_greens_2010}, \cref{eq:eq1} can be separated into the Hamiltonians of center of mass and relative degrees of freedom.
Since the $s$-wave interactions involve only the relative distance between a pair of particles, the center-of-mass Hamiltonian is fully decoupled, meaning that the relative one retains all the relevant information of the three-body system.
Therefore, we focus only on the relative Hamiltonian which gives, after transforming it into hyperspherical coordinates (for details see \cite{greene_review_2017}), the following expression:
\begin{equation}
    \label{eq:eq2}
    H_{\rm{rel}}=-\frac{\hbar^2}{2 \mu R^{5/2}} \frac{\partial^2}{\partial R^2}R^{5/2} +H_{\rm{ad}}(R;\Omega),
\end{equation}
where $\mu=\sqrt{m_1 m_2 m_3/(m_1+m_2+m_3)}\equiv m_H/\sqrt{1+2m_H/m_L}$ indicates the three-body reduced mass, $R$ is the hyperradius, and $\Omega$ is a collective coordinate denoting the five hyperangles \cite{avery_hyperspherical_1989,smirnov_method_1977}.
$H_{\rm{ad}}(R;\Omega)$ represents the part of the Hamiltonian which contains the hyperangular centrifugal potential as well as the two-body interactions expressed in the hyperspherical coordinates.
\begin{equation}
    \label{eq:eq3}
    H_{\rm{ad}}(R;\Omega)= \frac{\hbar^2}{2 \mu} \hat{\Lambda}^2 + \frac{15\hbar^2}{8\mu R^2} +\sum_{i>j}V_{ij}(R; \Omega), 
\end{equation}
where $\hat{\Lambda}$ denotes the grand angular momentum operator.

\begin{figure}[t!]
\centering
 \includegraphics[scale=0.40]{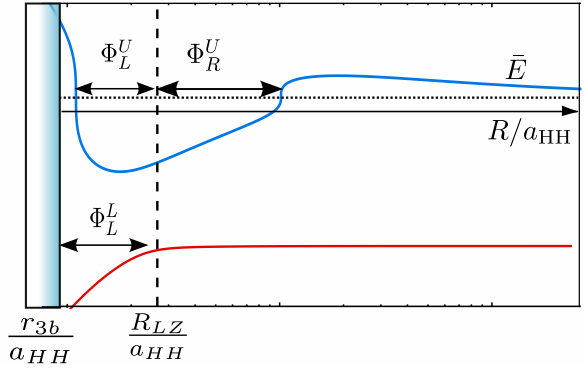}
 \caption{(Color on line) An illustration of the lowest hyperspherical potential curves $U_\nu^{1/3}(R/a_{HH})$ with $a_{HH}>0$ and $a_{HL}<0$. The red (blue) line saturates at large hyperradii in the atom+dimer (three-body break-up) threshold.
 The quantities $\Phi^U_L$, and $\Phi^U_L$ indicate the JWKB phase accumulation in the upper potential curve. 
 For the lower potential the corresponding phase is denoted by $\Phi^L_L$. The vertical dashed line represents the hyperradius where the non-adiabatic coupling $P-$matrix element $P_{12}$ maximizes. The horizontal dotted line refers to the three-body collisional energy $\bar{E}$ in units of $\frac{\hbar^2}{m_H a_{HH}^2}$, and the three-body parameter, $\frac{r_{3b}}{a_{HH}}$ depicted by the blue region.}
\label{fig:fig1}
\end{figure}

In the spirit of the adiabatic hyperspherical representation, the properly symmetrized three-body wave function is provided by the following ansatz:
\begin{equation}
    \Psi(R,\Omega)= \frac{1}{R^{5/2}}\sum_\nu \phi_\nu(R;\Omega) F_\nu(R),
    \label{eq:eq4}
\end{equation}
where $F_\nu(R)$ and $\phi_\nu(R;\Omega)$ indicate the $\nu$-th hyperradial and hyperangular part of the wave function, respectively.
In particular, $\phi_\nu(R;\Omega)$ components of $\Psi(R,\Omega)$ are obtained by diagonalizing \cref{eq:eq3} at a fixed hyperradius $R$.
\begin{equation}
  H_{\rm{ad}}(R;\Omega) \phi_\nu(R;\Omega)=U_\nu(R)\phi_\nu(R;\Omega),
  \label{eq:eq5}
\end{equation}
where the eigenvalues $U_\nu(R)$ are the so-called {\it adiabatic hyperspherical potential curves}.

Substitution of \cref{eq:eq4,eq:eq5} into the Schr\"odinger equation of the Hamiltonian $H_{\rm{rel}}$ and integration over all the hyperangles $\Omega$ yields a set of coupled ordinary second-order differential equations that solely depend on the hyperradius $R$.
\begin{equation}
  \bigg[-\frac{d^2}{d R^2}+\frac{2 \mu}{\hbar^2}(U_\nu(R)-E)\bigg]F_\nu(R)= \sum_{\nu '}V_{\nu \nu'}(R)F_{\nu '}(R),
  \label{eq:eq6}
\end{equation}
where $V_{\nu \nu'}(R)$ indicate the non-adiabatic coupling matrix elements/operators that are given by the following expressions:

\begin{subequations}
    \begin{align}
    &V_{\nu \nu'}(R)=2P_{\nu \nu '}(R)\frac{d}{dR}+ Q_{\nu \nu'}(R) ~~\rm{with}\label{eq:eq7a}\\
    &P_{\nu \nu'}(R)=\braket{\phi_\nu(R; \Omega)|\frac{\partial}{\partial R}\phi_{\nu '}(R;\Omega)}_{\Omega} \label{eq:eq7b}\\
    &Q_{\nu \nu'}(R)=\braket{\phi_\nu(R; \Omega)|\frac{\partial^2}{\partial R^2}\phi_{\nu '}(R;\Omega)}_{\Omega},\label{eq:eq7c}
    \end{align}
\end{subequations}
where $\braket{\ldots}_\Omega$ denotes that the integration over the hyperangles only.

Owing to the zero-range interactions, the non-adiabatic coupling matrix elements, $P_{\nu \nu'}(R)$ and $Q_{\nu \nu'}(R)$, as well as the hyperspherical potential curves $U_\nu(R)$ can be calculated semi-analytically, \cite{rittenhouse_greens_2010,kartavtsev_low-energy_2007,kartavtsev_universal_2006,nielsen_three-body_2001}.
However, the resulting hyperspherical potential curves $U_\nu(R)$, especially the lowest one, possess attractive singularities at the origin, i.e. the Thomas collapse. 
Therefore, an auxiliary parameter is introduced in order to truncate the attractive singularity in the potential curves, which in its simplest form consists of a hard wall placed at a small hyperradius, $R\approx r_{3b}$.
The three-body parameter $r_{3b}$ is arbitrary (from the point of view of zero-range theory) and is usually fixed via experimental observations.
In addition, the zero-range approximation greatly simplifies the computational cost since only hyperradial equations in \cref{eq:eq6} require numerical solution using standardized R-matrix methods \cite{aymar_multichannel_1996,mehta_three-body_2007,burke_jr_theoretical_1999}.

\cref{fig:fig1} depicts the two lowest hyperspherical potential curves $U^{1/3}_\nu(R/a_{HH})$ as obtained from zero-range approximation. 
The upper (blue) potential that vanishes at large hyperradii $R$ in the break-up threshold and the lower (red) potential which in the limit of large $R$ approaches the energy of the HH dimer. 
The light blue region denotes the hard wall boundary condition at $r_{3b}/a_{HH}$ that removes the attractive singularity of the lower curve.
The potential curves of \cref{fig:fig1} suffice in order to intuitively understand the recombination of three free particles into a universal pair of atoms with a recoiling one.
Consider the three-body system at a collisional energy $\bar{E}$ (in units of $\frac{\hbar^2}{m_H a_{HH}^2}$) indicated by the dotted line in \cref{fig:fig1}. 
In particular, we are interested in the low-energy limit in order to validate the two-channel approximation and highlight the threshold behavior of three-body collisions in HHL settings.
Viewing this three-body system heuristically as a time-dependent collision, starting from infinite long distances, the three-particles propagate inwards in the upper potential curve and tunnel with some probability under the repulsive barrier and then probe the corresponding classical allowed region at short hyperradii.
In this region, the non-adiabatic $P-$matrix element $P_{12}$ between the upper and lower potential curve plays a key role in inducing transitions. 
More specifically, at distances $R_{LZ}/a_{HH}$ (vertical dashed line) the corresponding $P$-matrix maximizes indicating the strong coupling regime.
This means that the particles transition with a certain probability from the upper to the lower curve and subsequently propagate outwards, fragmenting into a two-body molecule plus a spectator atom.
This recombination process is quantified mainly by evaluating the $|S_{12}|^2$ element of the scattering $S-$matrix.

As was shown in Ref.\cite{giannakeasprl2018}, the $|S_{12}|^2$ matrix element can be obtained analytically within the two-channel approximation by combining the Landau-Zener physics with the Jeffreys-Wentzel–Kramers–Brillouin (JWKB) approach.
The main constituents of this semi-classical approach are depicted in \cref{fig:fig1}.
More specifically, we assume that the $P-$matrix element of the potential curves in \cref{fig:fig1} possesses a Lorentzian lineshape in the vicinity or $R\approx R_{LZ}$, and we include the Langer correction in JWKB integrals \cite{nielsen_low-energy_1999}.
Under these considerations, the $|S_{12}|^2$ matrix element for the hyperspherical potential curves in \cref{fig:fig1} reads:
\begin{widetext}
\begin{equation}
    \label{eq:eq8}
    |S_{12}|^2=\frac{e^{-2\tau} p(1-p)\cos^2(\Phi^U_L-\Phi^L_L-\frac{\pi}{4}+\lambda)}{(1-\frac{e^{-4\tau}}{16})[p\cos^2(\Phi_L^L+\Phi^U_R-\frac{\pi}{4})+(1-p)\cos^2(\Phi_L^U+\Phi^U_R+\lambda)]-(1-\frac{e^{-2\tau}}{4})^2 p(1-p)\cos^2(\Phi^U_L-\Phi^L_L-\frac{\pi}{4}+\lambda)+\frac{e^{-4\tau}}{16}},
\end{equation}
\end{widetext}
where $e^{-2\tau}$ indicates the tunneling probability in a single collision with the repulsive barrier of the upper potential curve in \cref{fig:fig1}. 
\begin{figure}[t!]
\centering
 \includegraphics[scale=0.35]{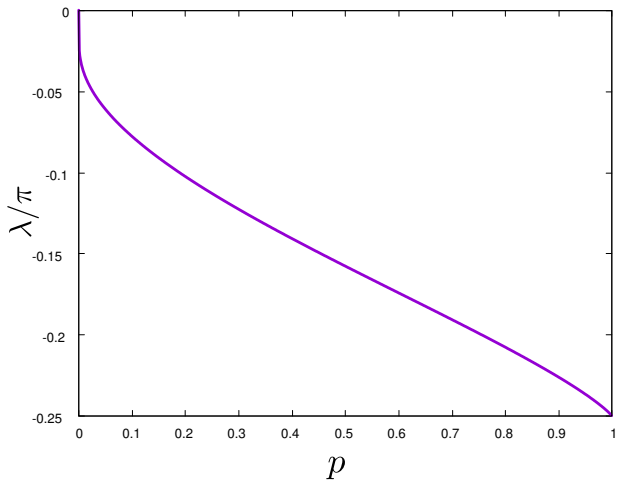}
 \caption{(Color on line) The Stokes correction phase as a function of the non-adiabatic probability $p$.}
\label{fig:fig2}
\end{figure}
The JWKB phases in the upper curve are indicated by the terms $\Phi^U_L$, and $\Phi^U_R$.
More specifically, $\Phi^U_L$ is the phase accumulation from the far left classical turning point up to $R\approx R_{LZ}$ whereas $\Phi^U_R$ is the JWKB integral from $R\approx R_{LZ}$ up to the inner classical turning point of the repulsive barrier.
Similarly, in the lower potential curve $\Phi^L_L$ corresponds to the phase accumulation between the hard wall (blue shaded region) located at $\frac{r_{3b}}{a_{HH}}$ and $R\approx R_{LZ}$.
Furthermore, $p$ corresponds to the Landau-Zener non-adiabatic probability to transition from the upper to the lower hyperspherical potential curve in a single pass through the avoided crossing region. 
The non-adiabatic probability $p$ is evaluated from the $P-$matrix elements which as we mentioned above are approximated to have a Lorentzian lineshape versus the hyperradius and a maximum at $R\approx R_{LZ}$ \cite{clark_calculation_1979}.
$\lambda$ is the Stokes phase and it is a correction added to the components of the hyperradial wave function, i.e. $F_\nu(R)$ with $\nu=1,2$, as they propagate through the non-adiabatic transition region \cite{child1974semiclassical,zhu1994theory}.
The Stokes correction phase depends on the non-adiabatic probability $p$ and it obeys the relation

\begin{equation}
    \label{eq:eq9}
    \lambda=\rm{arg}{\Gamma \bigg(i \frac{\delta}{\pi}\bigg)} -\frac{\delta}{\pi}\ln{\frac{\delta}{\pi}}+\frac{\delta}{\pi}+\frac{\pi}{4},
\end{equation}
where $\delta=-\frac{\ln{p}}{2}$. 
\cref{fig:fig2} shows the Stokes phase versus $p$ where in the diabatic (adiabatic) limit, i.e. $p=1~(0)$, the Stokes phase tends to $\lambda=-\pi/4~(0)$.
\cref{eq:eq8} captures the two main effects that occur in HHL systems. More specifically, the roots of the numerator of \cref{eq:eq8} indicate the St\"uckelberg suppression effects minimizing the probability of the HHL system to recombine into a shallow dimer.
On the other hand, the roots of the denominator in \cref{eq:eq8} denote the Efimov resonance phenomenon that enhances the recombination into a shallow dimer.
An additional insight obtained by \cref{eq:eq8} is that the St\"uckelberg suppression effects depend on the three-body parameter due to the phase $\Phi_L^L$ accumulation in the lower potential curve. In principle, also the Efimov resonances depend on $r_{3b}$, however, as \cref{eq:eq8} suggests in the limit of adiabatic collisions, i.e. $p\ll1$, only the phase accumulation in the upper potential curve survives which are independent of the three-body parameter meaning that such collisions possess a universal character.

\begin{figure}[b!]
\centering
 \includegraphics[scale=0.35]{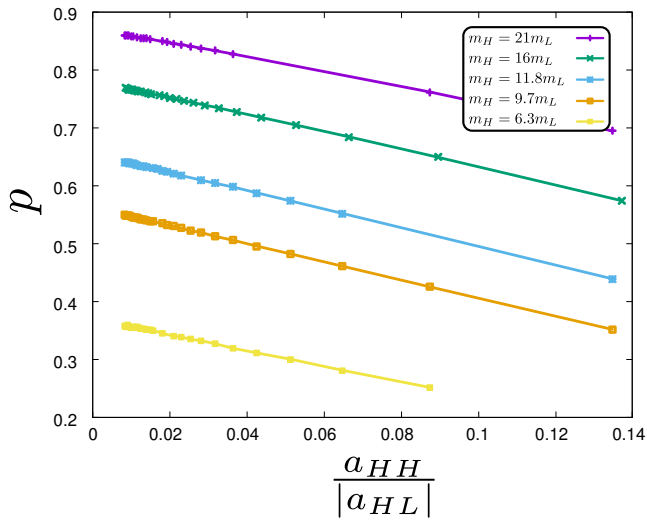}
 \caption{(Color on line) The degree of diabaticity $p$ as a function of the scattering length ratio $a_{HH}/|a_{HL}|$ for different mass ratios $m_H/m_L$ covering the regime of strong-to-weak mass-imbalanced three-body systems.}
\label{fig:fig3}
\end{figure}

The {\it degree of diabaticity} $p$ is depicted in \cref{fig:fig3}, in the zero-energy limit, as a function of $a_{HH}/|a_{HL}|$ for different mass ratios $m_H/m_L$ covering in this manner the regime from strong-to-weak mass-imbalanced atomic ensembles.
Note that we consider values of the ratio $a_{HH}/|a_{HL}|$ that correspond to $|a_{HL}|$ and $a_{HH}$ both being larger than the van der Waals length scales of the HL and HH pairs of atoms, respectively, ensuring the validity of the zero-range theory.
In particular, we observe in \cref{fig:fig3} that for large mass ratio, i.e. $m_H/m_L=21$, the corresponding three-body collision is more diabatic than in the case of weak mass-imbalance, i.e. $m_H/m_L=6.3$. 
This means that HHL systems with strong mass-imbalance, i.e. $m_H/m_L>21$, can easily transition from the three-body continuum to the shallow dimer+atom channel implying that the corresponding recombination process is strongly affected by the three-body parameter $r_{3b}$.
This behavior of the non-adiabatic probability $p$ on $m_H/m_L$ can be understood in terms of the ratio of the $P$-matrix elements and the energy difference of the hyperspherical potential curves, i.e. $\Delta$, at $R=R_{LZ}$.
According to Ref.\cite{clark_calculation_1979}, the probability $p$ is given by the relation $p=e^{-\pi \Delta/[4 v P_{12}(R_{LZ})]}$ where $v$ refers to the semi-classical velocity of the particles at $R=R_{LZ}$. 
Thus, for $a_{HL}\to -\infty$, the ratio of the energy gap $\Delta$ and $P_{12}(R_{LZ})$ increases as $m_H/m_L$ decreases yielding in return a decreasing probability $p$ and vice versa.

\begin{figure*}[t!]
\centering
 \includegraphics[scale=0.73]{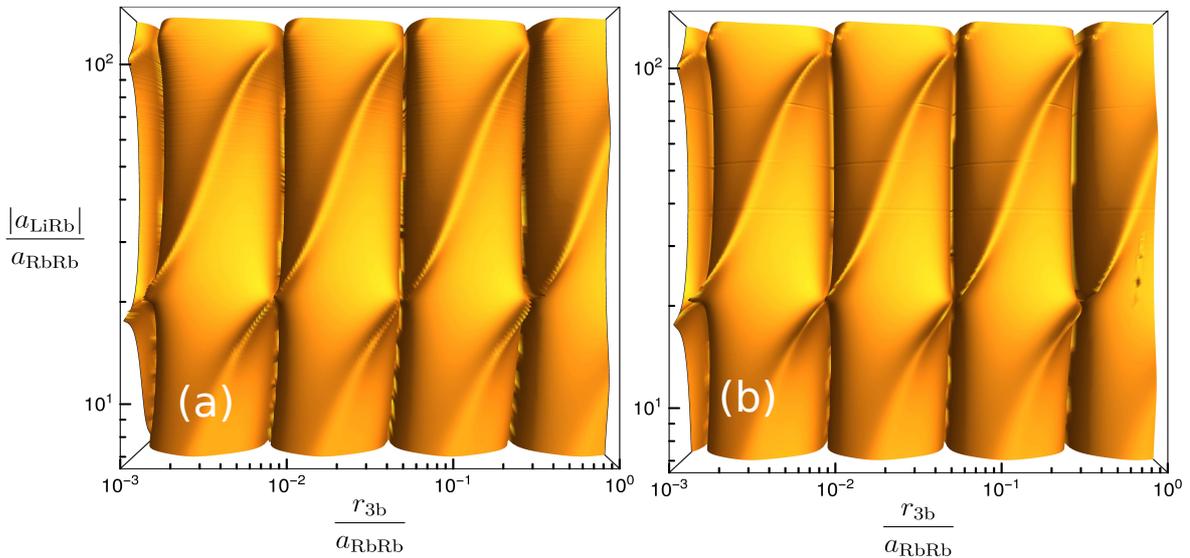}
 \caption{(Color on line) The scaled $\frac{|S_{12}|^2}{(k a_{HL})^4}$ matrix element versus the ratios $\frac{|a_{HL}|}{a_{HH}}$ and $\frac{r_{3b}}{a_{HH}}$ for the $^6\rm{Li}-^{87}\rm{Rb}-^{87}\rm{Rb}$ system at low-energy $E=\frac{\hbar^2 k^2}{2\mu}$. (a) semi-classical approach and (b) R-matrix numerical calculations.}
\label{fig:fig4}
\end{figure*}

As an example, \cref{fig:fig4} illustrates the scaled $S-$matrix element $\frac{|S_{12}|^2}{(k a_{HL})^4}$ for  $^6\rm{Li}-^{87}\rm{Rb}-^{87}\rm{Rb}$ three-body system at low-energies $E=\frac{\hbar^2 k^2}{2 \mu}$.
\cref{fig:fig4}(a) corresponds to the semi-classical model using \cref{eq:eq8} and \cref{fig:fig4}(b) refers to the case where the hyperradial equations are solved numerically within the R-matrix approach.
Both panels are in excellent agreement and the qualitative features, i.e. the enhancement and suppression of $\frac{|S_{12}|^2}{(k a_{HL})^4}$, are similar to those shown in Ref.\cite{giannakeasprl2018}.
In particular, as discussed in Ref.\cite{giannakeasprl2018} the enhancement of $\frac{|S_{12}|^2}{(k a_{HL})^4}$ is associated with an Efimov resonance. 
Namely, the upper potential curve in \cref{fig:fig1} can support a quasi-bound state behind the repulsive barrier at specific values of the ratios $\frac{|a_{HL}|}{a_{HH}}$ and $\frac{r_{3b}}{a_{HH}}$. 
Therefore, for colliding energies $E$ that match the energy of the quasi-bound three-body state, the atoms can easily tunnel under the barrier, where they can probe the non-adiabatic transition region and eventually hop with some probability to the $\rm{Rb}_2+\rm{Li}$ channel. 
Therefore, the presence of quasi-bound state in the upper potential curve in \cref{fig:fig1} causes $\frac{|S_{12}|^2}{(k a_{HL})^4}$ to be more pronounced.
On the other hand, the suppression of $\frac{|S_{12}|^2}{(k a_{HL})^4}$ is a manifestation of St\"uckelberg physics due to destructive interference of the alternative pathways, which prevents the three particles from exiting to infinity along the $\rm{Rb}_2+\rm{Li}$ channel.
However, in Ref.\cite{giannakeasprl2018} the $^6\rm{Li}-^{133}\rm{Cs}-^{133}\rm{Cs}$ system was investigated and the corresponding  $\frac{|S_{12}|^2}{(k a_{HL})^4}$ possesses one main qualitative difference from $^6\rm{Li}-^{87}\rm{Rb}-^{87}\rm{Rb}$.
Specifically, the $^6\rm{Li}-^{133}\rm{Cs}-^{133}\rm{Cs}$ system exhibits narrower Efimov resonances [see Fig.2(b) and (c) in Ref.\cite{giannakeasprl2018}] than those shown \cref{fig:fig4} for $^6\rm{Li}-^{87}\rm{Rb}-^{87}\rm{Rb}$.
This difference mainly arises from the fact that the collisions in $^6\rm{Li}-^{133}\rm{Cs}-^{133}\rm{Cs}$ are more diabatic than in $^6\rm{Li}-^{87}\rm{Rb}-^{87}\rm{Rb}$. 
As \cref{fig:fig3} suggests the non-adiabatic probability $p$ for $^6\rm{Li}-^{133}\rm{Cs}-^{133}\rm{Cs}$ is much closer to unit that for the case of $^6\rm{Li}-^{87}\rm{Rb}-^{87}\rm{Rb}$.
The lower values of $p$ for $^6\rm{Li}-^{87}\rm{Rb}-^{87}\rm{Rb}$ indicate the weak coupling of the quasi-bound Efimov state to the atom-dimer continuum which in return is manifested as a broad resonance in $\frac{|S_{12}|^2}{(k a_{HL})^4}$ matrix element.

\subsection{A simplified semi-classical model} \label{sec:sec1b}
In the following, we focus on the derivation of a simplified semi-classical model based on the prescription given in Ref.\cite{dincao_scattering_2005}.
Our goal is to unveil the scaling behavior of $S_{12}$ matrix element with respect to the length scales and the degree of diabaticity $p$ that govern HHL systems by incorporating only the necessary approximations.
Therefore, for our purposes from this point and on we assume that the colliding energy of the three atoms tends to zero, i.e. $E=\hbar^2 k^2/2\mu \to 0$.

\begin{figure}[t]
\centering
 \includegraphics[scale=0.40]{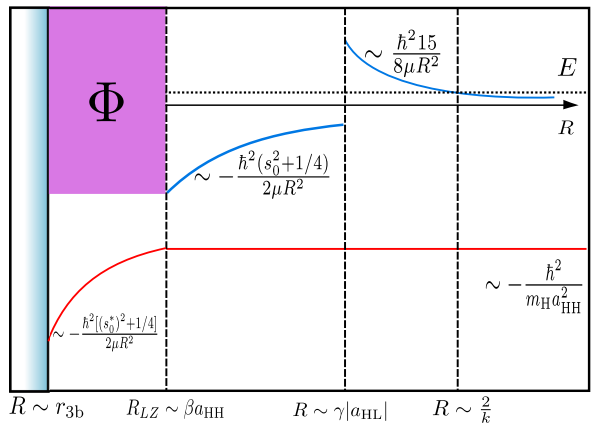}
 \caption{(Color online) An illustration of the approximate hyperspherical potential curves shown in \cref{fig:fig1} where $s_0$ and $s_0^*$ are the universal Efimov scaling coefficients. These piecewise curves are used in \cref{eq:eq10a,eq:eq10b,eq:eq11}.}
\label{fig:fig5}
\end{figure}

As in \cref{fig:fig1}, \cref{fig:fig5} illustrates the two-lowest hyperspherical potential curves which are properly parameterized using only the limiting tails of the curves of \cref{fig:fig1} in a piecewise manner.
Namely, for the upper potential curve (blue line) in \cref{fig:fig5} the universal tail $U_1(R)=-\frac{\hbar^2}{2\mu R^2}(s_0^2+1/4)$ is shown for hyperradii ranging from the non-adiabatic transition region, i.e. $R_{LZ}$, up to $R\sim \gamma |a_{HL}|$. 
Also, for $R>\gamma|a_{HL}|$ we consider only the tail of the repulsive barrier of the potential curve shown in \cref{fig:fig1} which falls off as $U_1(R)\sim \frac{\hbar^2}{2 \mu R^2}(15/4)$ with the outer classical turning point being located at $R\sim2/k$.
In addition, the effects of motion along the upper potential curve for $R<R_{LZ}$ is mapped to an arbitrary phase $\Phi$.
For the lower curve, at small hyperadii we employ the universal tail $U_2(R)=-\frac{\hbar^2}{2\mu R^2}[(s_0^*)^2+1/4]$ whereas for $R>R_{LZ}$ we assume that the potential curve is constant with energy equal to the heavy-heavy dimer.
Note that the parameters $s_0$ and $s_0^*$ correspond to the universal Efimov scaling coefficients for two and three resonant two-body interactions, respectively, and they are tabulated in Ref.\cite{wang_universal_2012-2} for several HHL systems.

Based on the piecewise potential curves of \cref{fig:fig5} and considering the low-energy limit, i.e. $k\to 0$, the tunneling amplitude $e^{-\tau}$, and the semi-classical phases $\Phi^U_L$, $\Phi^U_R$ and $\Phi^L_L$ are given by the following expressions:
\begin{subequations}
\begin{align}
    &e^{-\tau}\approx \int_{\gamma |a_{HL}|}^{2/k} \sqrt{4/R^2} \approx (\gamma k a_{HL}/2)^2,~~\Phi^U_L=\Phi,  \label{eq:eq10a}\\
    &\Phi^U_R \approx \int^{\gamma |a_{HL}|}_{\beta |a_{HL}|} \sqrt{s_0^2/R^2} \approx s_0 \ln \frac{\gamma |a_{HL}|}{\beta a_{HH}} \label{eq:eq10b}\\
    &~{\rm{and}}~\Phi^L_L\approx \int^{\beta |a_{HL}|}_{r_{3b}} \sqrt{(s_0^*)^2/R^2} \approx s_0^*\ln \frac{\beta a_{HH}}{r_{3b}},\label{eq:eq10c}
\end{align}
\end{subequations}

where the dimensionless parameters $\beta$ and $\gamma$ define the interval of hyperradius R such that the upper potential curve has the form $U_1(R)=-\frac{\hbar^2}{2\mu R^2}(s_0^2+1/4)$. In general, $\beta$ and $\gamma$ are considered as free parameters and they can be fixed by a fitting procedure to experimental or numerical data. Also, recall that the above JWKB integrals include the Langer corrections.
After substitution of \cref{eq:eq10a,eq:eq10b,eq:eq10c} into \cref{eq:eq8}, the $S-$matrix element $S_{12}$ reads

\begin{align}
        \label{eq:eq11}
            \frac{|S_{12}|^2}{(ka_{HL})^4 }&=\frac{\gamma^4}{16}p\cos^2\bigg(s_0^*\ln \frac{r_{\rm{3b}}}{a_{HH}}+\psi_1+\lambda\bigg)\bigg\{\frac{p}{1-p} \times \nonumber\\
        &\times\sin^2[s_0 \ln \frac{|a_{HL}|}{a_{HH}}+\psi_2-(s_0^*\ln \frac{r_{\rm{3b}}}{a_{HH}}+\psi_1)]\nonumber \\
        &+\cos^2(s_0 \ln \frac{|a_{HL}|}{a_{HH}}+\psi_2+\lambda) \nonumber \\
        &-p\cos^2(s_0^*\ln \frac{r_{\rm{3b}}}{a_{HH}}+\psi_1+\lambda)\bigg\}^{-1},
\end{align}
where the terms $(1-(\gamma k a_{HL}/2)^8/16)\approx1$, and $(1-(\gamma k a_{HL}/2)^4/4)\approx1$ since we focus on the low-energy regime, i.e. $k\to0$.
The phases $\psi_1$ and $\psi_2$ obey the expressions $\psi_1=\Phi-s_0^*\ln \beta-\pi/4$ and $\psi_2=\Phi+s_0\ln(\gamma/\beta)$, respectively.

\cref{eq:eq11} captures the main properties of the $S-$matrix element $S_{12}$ shown in \cref{fig:fig4}.
The numerator of \cref{eq:eq11} describes the positions of the St\"uckelberg interference minima which, as shown in \cref{fig:fig4},  scale logarithmically with respect to the ratio $r_{3b}/a_{HH}$. Also, the spacing between successive minima is constant on a logarithmic scale, and related to the universal Efimov scaling coefficient $s_0^*$.
On the other hand, the roots of the denominator of \cref{eq:eq11} trace out the maxima of $\frac{|S_{12}|^2}{(k a_{HL})^4}$ in \cref{fig:fig4}, i.e. the Efimov resonances, where the position of the successive resonances is defined by the $s_0$ universal factor.
We note \cref{eq:eq11} due to its simple structure can be used as a fitting formula for experimental measurements by treating the ($\psi_1$, $\psi_2$, $\gamma$) or ($\Phi$, $\beta$, $\gamma$)  as fitting parameters.

\section{Asymmetric lineshapes in three-body recombination coefficient} \label{sec:sec2}
\cref{fig:fig4} demonstrates that recombination resonant features are intertwined with St\"uckelberg interference minima.
This constitutes a unique feature of mass-imbalanced systems since for homonuclear three-body collisions the corresponding $S$- matrix element exhibits either Efimov resonances or St\"uckeleberg suppression effects for negative or positive scattering lengths, respectively.
Therefore, this section focuses on the lineshape of the $|S_{12}|^2$ squared matrix element plotted as a function of the ratio $\frac{r_{3b}}{a_{HH}}$ at fixed values of $\frac{|a_{HL}|}{a_{HH}}$.
In order to demonstrate the asymmetric lineshape of the Efimov resonances in HHL systems suffices to consider a range of $\frac{r_{3b}}{a_{HH}}$ values in the neighborhood of a St\"uckelberg minimum assuming a total colliding energy $E\approx 0$.

Under these considerations, utilizing the Fano profile formula \cref{eq:eq11} can be expressed in terms of the width of the resonance, $\Gamma$, and the Fano $q-$parameter which describes the asymmetry of the profile of the $|S_{12}|^2$.
\begin{align}
    &\frac{|S_{12}|^2}{(k a_{HL})^4}=A\frac{(x+ q)^2}{x^2+1},~~\rm{with} \label{eq:eq12} \\
    &A=\frac{\gamma^4 (1-p)}{16}\sin^2(s_0^*x_r+\psi_1+\lambda)\bigg\{\cos\big[2(s_0 \ln \frac{|a_{HL}|}{a_{HH}} \nonumber \\
    &-s_0^*x_r +\psi_2-\psi_1)\big] +(1-p)\cos\big[2(s_0^*x_r+\psi_1+\lambda)\big]\bigg\}^{-1},\nonumber
\end{align}
where $x=2(\ln \frac{r_{3b}}{a_{HH}}-x_r)/\Gamma$ with $x_r$ referring to the values of the ratio $\ln \frac{r_{3b}}{a_{HH}}$ that minimize the denominator of \cref{eq:eq11} at fixed $\frac{|a_{HL}|}{a_{HH}}$.
Note that \cref{eq:eq12} has same functional form as the conventional Fano formula, i.e. $\sigma=\sigma_0(\epsilon+q)^2/(\epsilon^2+1)$ \cite{fanophysrev1961}, where the ratio $\ln \frac{r_{3b}}{a_{HH}}$ is the independent variable instead of the energy.
For \cref{eq:eq12} the Fano lineshape asymmetry parameter $q$ and the width $\Gamma$ are given by the following expressions:
\begin{subequations}
\begin{align}
    q&= -\frac{2 }{s_0^*\Gamma}\cot(s_0^*x_r+\psi_1+\lambda) \label{eq:eq13a}~~{\rm{and}} \\
    \bigg(\frac{\Gamma}{2}\bigg)^2&=\frac{1-p}{(s_0^*)^2p}\bigg[\cos^2(s_0\ln \frac{|a_{HL}|}{a_{HH}}+\psi_2+\lambda)-p\cos^2(s_0^*x_r+ \nonumber \\
    &+\psi_1+\lambda) +\frac{p}{1-p}\sin^2(s_0\ln\frac{|a_{HL}|}{a_{HH}}-s_0^*x_r+\psi_2-\psi_1)\bigg]\times \nonumber \\
    &\times\bigg\{\cos[2(s_0 \ln \frac{|a_{HL}|}{a_{HH}} -s_0^*x_r+\psi_2-\psi_1)]+ \nonumber \\
    & +(1-p)\cos[2(s_0^*x_r+\psi_1+\lambda)]\bigg\}^{-1}. \label{eq:eq13b}
\end{align}
\end{subequations}
Note that $\Gamma$ is dimensionless here, in contrast to the usual Fano lineshape where $\Gamma$ has units of energy (or frequency).
\begin{figure}[t]
\centering
 \includegraphics[scale=0.19]{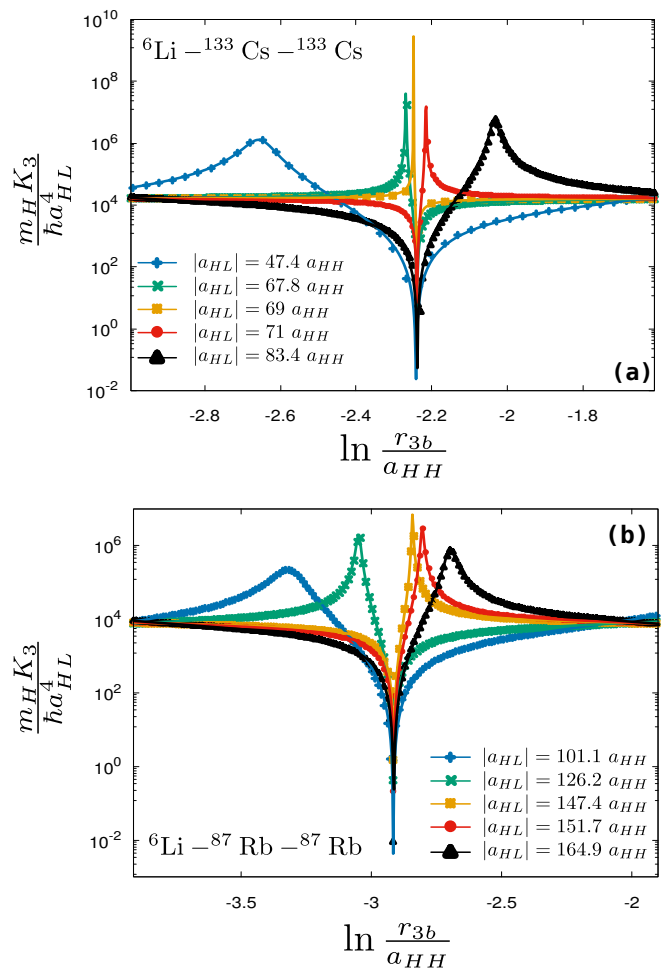}
 \caption{(Color online) In the limit of $E\to0$, the scaled recombination coefficient $\frac{m_H K_3}{\hbar a_{HL}^4}$ is shown as a function of  $\ln \frac{r_{3b}}{a_{HH}}$ for (a) $^6\rm{Li}-^{133}\rm{Cs}-^{133}\rm{Cs}$ and (b) $^6\rm{Li}-^{87}\rm{Rb}-^{87}\rm{Rb}$. The symbols refer to the corresponding calculations in the semi-classical approach. The solid lines indicate the fitting of \cref{eq:eq11} using the universal parameters shown in \cref{tab:tab1}.}
\label{fig:fig6}
\end{figure}
The three-body recombination coefficient of HHL systems can be expressed in terms of the $S_{12}$ matrix element yielding the relation
\begin{equation}
    \label{eq:eq13}
    K_3=\frac{64 \hbar \pi^2}{\mu k^4} |S_{12}|^2,
\end{equation}
where $k=\sqrt{2\mu E/\hbar^2}$ with $E$ being the total colliding energy of the three-body system. 

\begin{table}[b!]
	\centering
\begin{tabular}{cccccccc}
 HHL System  & $s_0$ & $s_0^*$ & $\gamma$ & $\psi_1$ & $\psi_2$ &\\
 \hline
 \\
$^{6}\rm{Li}-^{133}Cs-^{133}Cs$ & 1.983 & 2.003 & 4.42 & 0.46 & 0.13&\\
\\
$^{6}\rm{Li}-^{87}Rb-^{87}Rb$ & 1.633 & 1.682 & 3.13 & 0.8 & 0.4&
\end{tabular}
\caption{ A summary of the universal parameters used in \cref{eq:eq12,eq:eq13a,eq:eq13b} for the systems of $^{6}\rm{Li}-^{133}Cs-^{133}Cs$ and $^{6}\rm{Li}-^{87}Rb-^{87}Rb$. Note that the values of $s_0$ and $s_0^*$ are calculated in Ref.\cite{wang_universal_2012-2}.}
	\label{tab:tab1}
\end{table}
\begin{figure}[t]
\centering
 \includegraphics[scale=0.25]{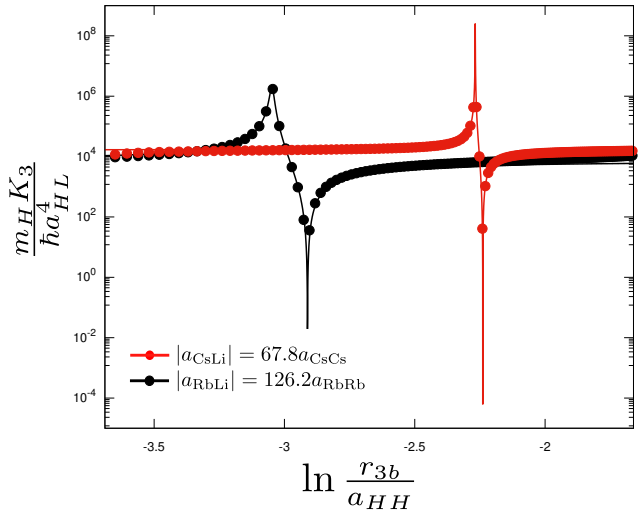}
 \caption{(Color online) A comparison of the scaled recombination coefficient obtained via the fitting of \cref{eq:eq11} (points) and the Fano lineshape formula (solid lines) from \cref{eq:eq12} for two HHL systems. The red points and lines correspond to $^{6}\rm{Li}-^{133}Cs-^{133}Cs$ for a scattering length ratio $\frac{|a_{\rm{CsLi}}|}{a_{\rm{CsCs}}}=67.8$. The black points and lines denote the $^{6}\rm{Li}-^{87}Rb-^{87}Rb$ system at $\frac{|a_{\rm{RbLi}}|}{a_{\rm{RbRb}}}=126.2$. Note that the total colliding energy is set to zero. 
 }
\label{fig:fig7}
\end{figure}

For a total colliding energy $E\approx 0$, \cref{fig:fig6}(a) and (b) depict the scaled recombination coefficient $\frac{m_H K_3}{\hbar a_{HL}^4}$ versus the ratio $\ln \frac{r_{3b}}{a_{HH}}$ in the vicinity of a St\"uckelberg minimum for two three-body systems, i.e. $^6\rm{Li}-^{133}\rm{Cs}-^{133}\rm{Cs}$ and $^6\rm{Li}-^{87}\rm{Rb}-^{87}\rm{Rb}$, respectively.
More specifically, the symbols in both panels correspond to the full semi-classical calculations whereas the solid lines are obtained by \cref{eq:eq11}, i.e. the simplified semi-classical model, using $\gamma$, $\psi_1$ and $\psi_2$ as fitting parameters.

Note that \cref{tab:tab1} summarizes the values of these parameters for both HHL systems exhibiting universal characteristics since they are independent of scattering length ratio $\frac{|a_{HL}|}{a_{HH}}$. 
Therefore, in order to extract the values of the $\gamma$, $\psi_1$ and $\psi_2$ parameters suffices to fit only the semiclassical calculations for $|a_{HL}|/a_{HH}=47.4$ and $|a_{HL}|/a_{HH}=101.1$ in panels (a) and (b), respectively.
However, the phases $\psi_1$ and $\psi_2$ and the amplitude $\gamma$ depend on the mass ratio of the HHL system since the corresponding hyperspherical potential curves are strongly influenced by variations of $m_H/m_L$.
Evidently, both panels showcase the asymmetric profile of the Efimov resonance as a distinctive feature of HHL systems where \cref{eq:eq11} is in excellent agreement with the corresponding semi-classical calculations.
In particular, in \cref{fig:fig6}(a) we observe that for scattering length ratios in the range $55<\frac{|a_{HL}|}{a_{HH}}<69$, the Efimov resonance occurs to the left of St\"uckelberg minimum and its width decreases towards $\frac{|a_{HL}|}{a_{HH}} \to 70$.
For $\frac{|a_{HL}|}{a_{HH}} > 70$, the Efimov resonance emerges to the right of the St\"uckelberg minimum with an increasing width.
This behavior of the resonant structure as a function of the ratio $\frac{|a_{HL}|}{a_{HH}}$ is known as the $q$-reversal effect, where the asymmetry parameter $q$ changes sign at $\frac{|a_{HL}|}{a_{HH}}\sim70$.
For $^6\rm{Li}-^{87}\rm{Rb}-^{87}\rm{Rb}$ shown in \cref{fig:fig6}(b) a similar behavior is observed demonstrating that the occurrence of $q$-reversal is independent of the particles' mass ratio.   
The $q$-reversal phenomenon is a manifestation of quantum interference and in HHL systems it occurs when $s_0^*x_r+\psi_1+\lambda= n \pi/2$ with $n$ being an integer.

Additionally, \cref{fig:fig7} demonstrates the validity of the Fano lineshape formula given in \cref{eq:eq12}.
More specifically, \cref{fig:fig7} illustrates a comparison of the scaled recombination coefficient between the fitting of \cref{eq:eq11} (red and black dots) and the Fano lineshape formula from \cref{eq:eq12} (red and black solid lines) at low collisional energies.
In particular, the red (black) symbols and lines refer to the $^6\rm{Li}-^{133}\rm{Cs}-^{133}\rm{Cs}$ ($^6\rm{Li}-^{87}\rm{Rb}-^{87}\rm{Rb}$) for a scattering length ratio $\frac{|a_{\rm{CsLi}}|}{a_{\rm{CsCs}}}=67.8$ ($\frac{|a_{\rm{RbLi}}|}{a_{\rm{RbRb}}}=126.2$).
We observe that the Fano lineshape formula from \cref{eq:eq12} is in good agreement with the corresponding semi-classical calculations of \cref{eq:eq11}.

\begin{figure}[t]
\centering
 \includegraphics[scale=0.1]{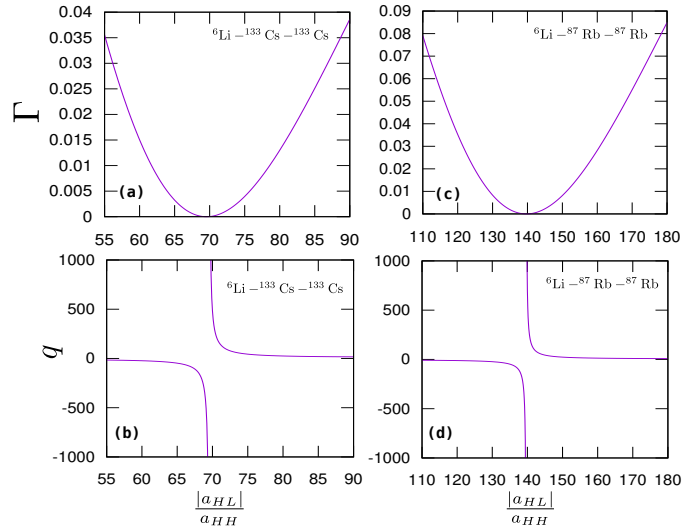}
 \caption{(Color online) Panels (a) [(c)] and (b) [(d)] show the width of the Efimov resonance $\Gamma$ and the asymmetry parameter $q$ versus the scattering length ratio $\frac{|a_{HL}|}{a_{HH}}$ for the $^6\rm{Li}-^{133}\rm{Cs}-^{133}\rm{Cs}$ ($^6\rm{Li}-^{87}\rm{Rb}-^{87}\rm{Rb}$) system, respectively. Note that the total colliding energy is set to zero. Also, $\Gamma$ and $q$ are obtained via \cref{eq:eq13a,eq:eq13b}, respectively, using the universal parameters shown in \cref{tab:tab1}. }
\label{fig:fig8}
\end{figure}

The width $\Gamma$ of the Efimov resonances and the lineshape asymmetry $q$ are shown in \cref{fig:fig8} for $^6\rm{Li}-^{133}\rm{Cs}-^{133}\rm{Cs}$ [see panels (a) and (b)] and $^6\rm{Li}-^{87}\rm{Rb}-^{87}\rm{Rb}$ [see panels (c) and (d)].
$\Gamma$ and $q$ are obtained via \cref{eq:eq13a,eq:eq13b} using the universal parameters of \cref{tab:tab1}.
In panels (b) and (d), we observe the $q-$reversal effect where at $\frac{|a_{HL}|}{a_{HH}}=70$ and $\frac{|a_{HL}|}{a_{HH}}=140$ the lineshape asymmetry $q$ diverges.
This implies that for large $q$ parameters the recombination coefficient approaches a symmetric lineshape that is centered at $x_r$.
Furthermore, we observe that at $|q|\to\infty$ the corresponding widths of the Efimov resonances tend to zero, i.e. $\Gamma \to 0$, as is illustrated in \cref{fig:fig8}(a) for $^6\rm{Li}-^{133}\rm{Cs}-^{133}\rm{Cs}$ and \cref{fig:fig8}(c) for $^6\rm{Li}-^{87}\rm{Rb}-^{87}\rm{Rb}$.
This means that in this range of parameters the Efimovian quasi-bound state stabilizes into a bound one which is fully decoupled from the three-body and the atom-dimer continua.
This counter-intuitive phenomenon is known as bound state in the continuum and such states have been observed in various fields of physics \cite{hsuBoundStatesContinuum2016}.

\section{Summary}
In summary, the properties of three-body recombination processes into shallow dimers for HHL systems are investigated.
Focusing on the low-energy regime, we consider inter- and intraspecies interactions that possess negative and positive scattering lengths, respectively, {thereby highlighting the threshold behavior of such HHL systems}.
For this three-body system, we have reviewed the theoretical methods used in Ref.\cite{giannakeasprl2018} and in particular, the semi-classical approach providing additional details on the Stokes phase and the degree of diabaticity $p$.
Furthermore, a simplified version of the semi-classical method is derived by approximating the hyperspherical curves with piecewise potential tails as in Ref.\cite{dincao_scattering_2005}.
The simplified semi-classical model provides closed form expressions of the $S-$matrix elements which describe the process of three free-particles recombining into the shallow dimer+atom channel.
Namely, we show that \cref{eq:eq11} captures all the main attributes of the recombination spectra for HHL systems, such as the asymmetric lineshape in the three-body recombination coefficient, the logarithmic scaling of the Efimov resonances and St\"uckelberg interference minima.
In particular, \cref{fig:fig6} demonstrates that \cref{eq:eq11} can be used as a fitting formula for the recombination spectra in HHL systems since the parameters $\psi_1$, $\psi_2$ and $\gamma$ are insensitive on the scattering length ratio $\frac{|a_{HL}|}{a_{HH}}$.
Focusing on the resonant profile of the recombination coefficient, \cref{eq:eq11} is parameterized in terms of the width of the resonance $\Gamma$ and the lineshape asymmetry $q$.
This parameterization enables us to identify two emergent phenomena that occur only in heteronuclear three-body collisions:  (i) the $q$-reversal effect which describes the change in the asymmetry of the profile of the three-body recombination coefficient as a function of the scattering length ratio $\frac{|a_{HL}|}{a_{HH}}$ and (ii) the modification of an Efimov resonance into a bound state embedded in the three-body and atom-dimer continua for $|q|\to \infty$. 

\section{Acknowledgment}
The authors would like to thank  M.T. Eiles for fruitful discussions.
The work of CHG was supported in part by the U.S. National Science Foundation, Grant No. PHY-1912350.
The numerical calculations were performed using NSF XSEDE Resource Allocation No. TG-PHY150003.

\bibliography{few_body.bib}
\end{document}